\documentclass[aps,prd,superscriptaddress,floatfix]{revtex4-1}
\usepackage{axodraw2}
\usepackage{color}
\usepackage{amssymb}
\usepackage[normalem]{ulem}
\usepackage{graphicx,color}
\usepackage{amsmath}
\usepackage{subfigure}
\usepackage[normalem]{ulem}
\usepackage{booktabs}
\usepackage{xcolor}
\usepackage{epsfig,graphics,color,graphicx,amsmath}
\usepackage{epstopdf}
\usepackage{epsfig,graphics,color,graphicx,amsmath}
\def\be{\begin{equation}}
	\def\ee{\end{equation}}
\def\bea{\begin{eqnarray}}
	\def\eea{\end{eqnarray}}

\colorlet{darkgreen}{green!50!black}
\colorlet{brightyellow}{yellow!75!red}
\colorlet{orange}{red!50!yellow}
\colorlet{darkblue}{blue!60!black}
\colorlet{darkred}{red!80!black}
\usepackage{array}
\usepackage{array,etoolbox}
\usepackage{fancyhdr}
\usepackage{slashed}

\newcommand{\bno}{\begin{eqnarray*}}
	\newcommand{\eno}{\end{eqnarray*}}

\newcommand{\bl}{\begin{large}}
	\newcommand{\el}{\end{large}}
\newcommand{\bla}{\begin{Large}}
	\newcommand{\ela}{\end{Large}}

\newcommand{\ede}{{\end{document}}}

\def\be{\begin{equation}}
\def\ee{\end{equation}}
\def\bea{\begin{eqnarray}}
\def\eea{\end{eqnarray}}
\def\psla{ \rlap \slash \! }  %% \!\!\!\!\!
\begin{document}
	 \vspace{-12ex} 
	%%	\vspace{-5ex}   
	\begin{flushright} 	
		{\normalsize \bf \hspace{3ex} LFTC-23-07/80 }
	\end{flushright}
	\title{ Pseudoscalar current  and covariance with the light-front approach}
	\author{ Jurandi Le\~ao}
	\affiliation{\em Laborat\'orio de F\'\i sica Te\'orica e Computacional-LFTC, 
		Universidade Cruzeiro do Sul / Universidade Cidade de S\~ao Paulo,
		015060-000, S\~ao Paulo, SP, Brazil 
	 and 
	Instituto Federal de S\~ao Paulo\\
	Avenida Bahia, Caraguatatuba \\ 
	11665-071 S\~ao Paulo,
	Brazil 
}
	\author{ J. P. B. C. de Melo}
	\affiliation{\em Laborat\'orio de F\'\i sica Te\'orica e Computacional-LFTC, 
		Universidade Cruzeiro do Sul / Universidade Cidade de S\~ao Paulo,
		015060-000, S\~ao Paulo, SP, Brazil}
\begin{abstract}
	Quantum Field Theory (QFT) is used to describe the physics of particles in terms of
	 their fundamental constituents. The Light-Front Field Theory~(LFFT), introduced 
	 by Paul Dirac in 1949~\cite{Dirac1949}, is an alternative approach to solve some of the problems that 
	 arise in quantum field theory.  The LFFT is similar to the Equal Time Quantum Field Theory~(EQT), 
	 however, some particularities  are not, such as the loss of covariance in the light-front.
	 	Pion electromagnetic form factor is studied in this work at lower and higher
	 	 momentum transfer regions to explore the constituent quark models and the differences among these and other models. 
	The electromagnetic current is calculated with both the ``plus'' and ``minus''
	 components in the light-front approach.	The results are compared with other
	  models, as well as with experimental data. 
\\
\newline
keys words: pion,light-front, quark model, electromagnetic current, 
electromagnetic form factor 
%PACS numbers:  \pacs{03.75.Fi, 32.80.Pj, 42.50.Md, 42.81.Dp}
\end{abstract}
\date{\today}   
\maketitle

\section{Introduction}   

Quantum chromodynamics~(QCD), the theory of strong interactions,  
consists in the study of one of the four fundamental interactions in the standard model. 
One of the most important questions in QCD, not 
yet resolved, is in regard of the non-perturbative regime. 
To address this particular issue is not straightforward, although lattice QCD has
been progressing and thus promising advances are expected.

Even with some relativistic constituent quark models, 
it is possible to study hadron physics in non-perturbative regions 
in an efficient manner with the quark and gluon degrees of freedom~\cite{Brodsky1998}. 
Before the advent of QCD, the pion, the lightest mass in hadrons regarded as the quark-antiquark 
bound state, has provided the long-range attractive part of the nucleon-nucleon
 interaction~\cite{Lacombe2002}. 
Amongst other approaches, the light-front aims to consistently
describe  the pion (hadron) bound state 
involving both the higher and lower momentum transfer regions.  
Hence, the light-front quantization has been used to 
compute the hadronic bound state wave functions~\cite{Brodsky1998,Harindranath2000}.
The advantage of the light-front approach compared to the instant form 
is its simplicity~\cite{Zuber}. %%%%,Weinberg}. 

In the light-front approach, the bound state wave functions are
defined in the hypersurface, $x^+=x^0+x^3=0$, and are covariant under
kinematical front-form boosts, because of the Fock-state decomposition stability~\cite{Perry1990}. 
Due to the simpleness in handling the wave functions and calculating observables,  
the light-front constituent quark model (LFCQM) has received a lot of 
attention in the past~\cite{Terentev1976,deAraujo1995}.
The LFCQMs achieved impressive success in describing the 
electromagnetic properties of hadrons, in particular 
those of the pseudoscalar and spin-1/2 
particles~\cite{Dziembowski1987,Cardarelli1995,Cardarelli1996,Krutov1998,deMelo1999,Krutov2009,Melikhov2002,deMelo2004,Cheng2004,Huang2004,Braguta2004,Salcedo2004v1,Salcedo2006,Karmanov2007,deMelo2002,Bakker2001,Kisslinger2001,deMelo2006,Krutov2009,Biernat2014,Yabusaki2015,Horn2016,Adhikari2016}, as well as spin-1 vector 
particles~\cite{deMelo1997,deMelo1999v2,Lev1999,Lev2000,Jaus2003,deMelo2003,Aliev2004,Braguta2004v2}.

The extraction of the electromagnetic form factor with the light-front approach 
depends on the electromagnetic current's component adopted, due to problems related with the rotational symmetry breaking and the zero modes, namely, a non-valence contribution to the electromagnetic current's matrix elements~\cite{deMelo1997,deMelo1999v2,Naus1998,Bakker2003}. 

It is discussed in \cite{deMelo1997,deMelo1999v2,Choi2004,Bakker2002,deMelo2012,Clayton2015}
 that, for spin-1 particles, the plus component of the electromagnetic current,~("$J^{+}$"), 
is not free from the pair term contributions~(or non-valence contributions) within the Breit frame 
($q^+=0$), and, thus, that the rotational symmetry is broken.

In general, the matrix elements of the electromagnetic 
current in the light-front formalism
have other contributions besides the valence contribution, 
namely the pair terms~\cite{deMelo2002,deMelo1999v2,Naus1998,deMelo2012}, 
in which the covariance is restored.
If the pair term contribution is taken correctly, it does not matter
which component of the electromagnetic current is used  
to extract the electromagnetic form factors of the hadrons. 
In the present work, two types of the vertex functions
for the $\pi-q\bar{q}$ vertex are assumed to calculate the pion
electromagnetic form-factor,   
and results for the both cases are compared with 
experimental data~\cite{Amendolia1984,Amendolia1986,Baldini1999,Volmer2001,Blok2002,Horn2006,Tadevosyan2007}. 
In the lower momentum transfer region, non-perturbative regime of QCD is more important than 
the perturbative regime, the latter working better in the higher momentum transfer region.

Studies of light-vector and pseudoscalar mesons 
in the light-front approach are very important, 
since they can provide a hint of the non-perturbative regime of QCD, 
and those involving the light-pseudoscalar mesons can shed light on 
the (spontaneous) chiral symmetry breaking. 
Properties of such mesons are also studied with other 
approaches 
\cite{Krutov2009,Roberts1996,Hawes1999,Maris2002,Aliev2004,
	Carvalho2004,Desplanques2005,Desplanques2009,Noguera2007,Santopinto2005,Giannini2005,
	Tomasi2005v1,Tomasi2005v2,Nesterenko1982,Braguta2004,Braguta2004v2,
	Gutierrez-Guerrero:2010waf,Chang2013nia,Raya:2015gva,Raya2019dnh}, 
as well as with the lattice formulation in the light-front~\cite{Dalley2003}.

For the lightest pseudoscalar meson, pion, the models based on 
the Schwinger-Dyson approach~\cite{Roberts1996,Maris2002} describe the 
electromagnetic form factor quite well.
However, some differences amongst the models including the light-front 
approach can be noticed. 
Here, we take the light-front models for the pion, presented
 in previous works~\cite{deMelo1999,deMelo2002}, and extend it to 
study the higher momentum transfer region and compare with other 
approaches such as the vector meson dominance~\cite{Kroll1967,Krein1993}. 

This paper is organized as follows. In Section II, the model of the wave function 
for the pion, the quark-antiquark bound state in the light-front, is presented,
and the electromagnetic form factor is calculated for both 
non-symmetric and symmetric $\pi-q\bar{q}$ vertices. 
In particular, for the non-symmetric vertex, the plus and minus 
components of the electromagnetic current are used, and the corresponding two 
results are shown, including the weak decay constant of the pion.
In Section III, the vector dominance model is introduced and compared with the light-front approach used. 
In Section IV, numerical results and discussions
 are presented and, finally, the conclusions are given in Section V.

\section{Light-Front Wave Function and Electromagnetic Current} %%Form Factor}

One of the main goals in the light-front approach is to solve the bound state equation,
\begin{equation}
H_{LF}| \Psi > = M^2| \Psi>, 
\label{auto}
\end{equation}  
where $H_{LF}$ is the light-front Hamiltonian and $M^2$ is the invariant mass  
associated with the physical particle, the eigenstate of the 
the light-front Hamiltonian~\cite{Brodsky1998}. 
The light-front wave function relates to the Bethe-Salpeter 
wave function~(see~\cite{deMelo2002} for more details). 
With the light-front wave function it is possible to calculate the matrix elements 
of the bound states. The meson state may be expressed by a superposition of all Fock states:

\begin{equation}
|\Psi_{meson}>=
\Psi_{q\bar{q}} |q\bar{q}>  +
\Psi_{q\bar{q}g}|q\bar{q}g>  +  \cdots .
\label{eq2}
\end{equation}
The electromagnetic current where $J^{\mu}$, is expressed in 
terms of the quark fields and the charge, i.e.,
~$J^{\mu}=\sum_{f} e_{f} \bar{q}_{f} 
\gamma_\mu q_{f}$, with $e_f$ and $q_f$ being, respectively, the charge and 
quark field of the flavor $f$ quark. In this way, the matrix elements of
the electromagnetic current are given by  
% \begin{multline}
\begin{equation}
 J^\mu  = -\imath 2 e \frac{m^2}{f^2_\pi}
 N_c\int \frac{d^4k}{(2\pi)^4} Tr \Bigl[ S(k)
 \gamma^5 S(k-p^{\prime})
 \gamma^\mu S(k-p) \gamma^5 \Bigr] 
\times \Gamma(k,p^{\prime})
 \Gamma(k,p) \ , 
 \label{jmu}
\end{equation}
% \end{multline}
where 
$\displaystyle S(p)=\frac{1}{\rlap\slash p-m+\imath \epsilon}$ 
is the quark propagator, $N_c=3$ is the number of colors and $m$ is the constituent quark mass. 
The factor 2 appears from the isospin algebra~\cite{deMelo1999,deMelo2002}.

The pion quark-antiquark vertex is constrained by the pseudoscalar attributes
	 for the pion \cite{Horn2016,Chang2013nia,Raya:2015gva,Roberts:2011wy}.
		In the presente work, we use the similar approach developed some years ago 
	by Frederico and Miller~\cite{Frederico1992ye},  
	where a effective Lagrangian for the pion quark vertex was utilized, 
	%% 	$${\cal L}_i=-\frac{\hat{m}{f_\pi}} \bf {pi} \bar{q} \gamma^5 \tau q $$. 
	%%	where only the $\bar{q}q$, 
		considering only the most important component of the vertex function for the 
		Bethe-Salpeter amplitude, i.e, proportional to  $\gamma^5$.  
		This effective Lagrangian it is associated to the meson vertex, 
		in order to build the spin and flavour structure of the pion meson.

The function $\Gamma(k,p)$ in Eq.~(\ref{jmu}) is the regulator vertex function 
to regularize the Feynman amplitude, i.e., the triangle diagram for the electromagnetic 
current. Here, we use two possible~$\pi-q\bar{q}$ vertex functions. The first one is the
non-symmetric vertex, used in the previous work~\cite{deMelo1999, daSilva2012},
\begin{equation}
\Gamma^{(NSY)}(k,p)=
\biggl[
\frac{N}{((p-k)^2-m^2_R+\imath\epsilon)}
\biggr]
\label{nosymm} \ ,
\end{equation} 
while the second one is a symmetric vertex, 
used in \cite{deMelo2002,Yabusaki2015}:
\begin{equation}  
\Gamma^{(SY)}(k,p)=
\biggl[ 
\frac{N}{(k^2-m^2_R + \imath\epsilon)} +
\frac{N}{((p-k)^2-m^2_R + \imath\epsilon)}
\biggr].
\label{symm}
\end{equation}
In equation the above, $m_R$ is the regulator mass used to keep 
the amplitudes finite, and represents the soft effects at the short range. 
An important property in the electromagnetic processes is the 
current conservation, or the gauge invariance. 
 The calculation is performed within the Breit frame;  the initial and 
 final momenta of the bound system are, respectively, $p^{\mu} = (p_0/2,-q/2 \cos \alpha, 0, -q/2 \sin \alpha)$
 and $p^{\prime {\mu}} = (p_0/2,q/2 \cos \alpha, 0, q/2 \sin \alpha)$. The transferred momentum is 
 $q^{\mu}=(0, q \cos \alpha, 0, q \sin \alpha)$ 
 and the spectator quark momentum is $k^{\mu}$.

The current conservation must be satisfied with the inclusion of
the two vertex functions, $\Gamma(k,p)$, as used in the present work. 
This conservation is indeed easily proved within Breit frame~(see Ref.~\cite{Naus1998} for the proof).

%% \begin{widetext}
%% Fig. figure by Axodraw
\hspace{-1.5cm}
\begin{figure}[t]
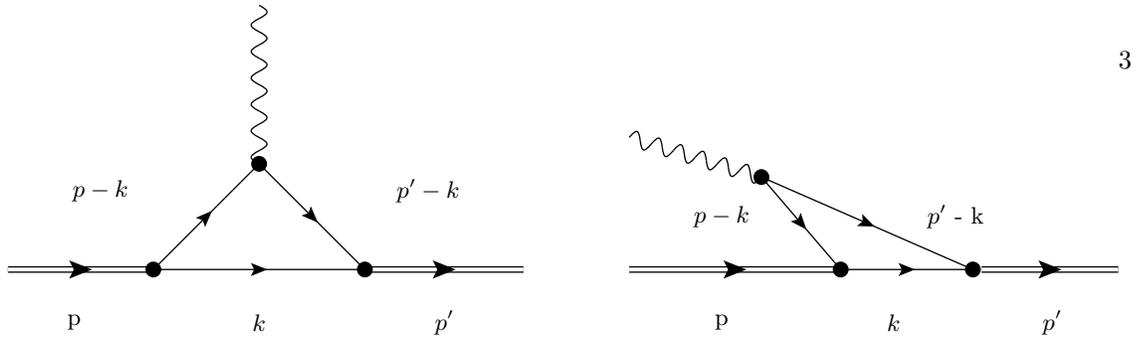

	\begin{center}
		\begin{axopicture}(350,110)
    	%% \SetScale{0.8}
			\SetScale{1.0} %%0.60}
			%% \SetColor{Red}
			\Photon(100,150)(100,90){3}{6}
			\Vertex(100,90){3.0}
			\Line[arrow](60,50)(100,90)
			\Line[arrow](100,90)(140,50)
			\Line[arrow,arrowpos=0.5,double,sep=2](5,50)(60,50)
			\Line[arrow](60,50)(140,50)
			\Text(30,30){p}
			\Text(40,80){$p-k$}
			\Text(164,80){$p^{\prime}-k$}
			\Text(100,30){$k$}
			\Text(170,30){$p^\prime$}
			%% \SetColor{Black}
			\Vertex(60,50){3.0}
			\Line[arrow,arrowpos=0.5,double,sep=2](140,50)(200,50)
			\Vertex(140,50){3.0}
			%% Second Feynman Diagram
			%%
			\Line[arrow,arrowpos=0.5,double,sep=2](240,50)(320,50)
			\Vertex(320,50){3.0}
			\Photon(240,100)(290,85){3}{6}  %% [arrowpos=0.5]{3}{6}
			\Vertex(290,85){3.0}
			\Line[arrow](290,85)(320,50)
			\Line[arrow](290,85)(370,50)
			\Line[arrow](320,50)(370,50)
			\Vertex(370,50){3.0}
			\Line[arrow,arrowpos=0.5,double,sep=2](373.3,50)(425,50)
			\Text(275,30){p}  
			\Text(275,70){$p - k$}
			\Text(365,70){$p^\prime$ -~k }
			\Text(340,30){$k$}
			\Text(400,30){$p^\prime$}
		\end{axopicture}
	\caption{Feynman diagrams for the valence contribution~(left panel) and the non-valence 
	contribuition~(right panel) for the electromagnetic current.} 
	\end{center}
\end{figure}
%%   \end{widetext}
%% 

The plus component of the electromagnetic current,~$J^{+}$, is used to
extract the pion electromagnetic form factor, where the Dirac ``plus" matrix is given by   
$\gamma^+=\gamma^0+\gamma^3$. The plus component ($J^{+}_{\pi} =J^0+J^3 $)
of the electromagnetic current 
for the pion, calculated in the light-front formalism through the triangle 
Feynman diagram in the impulse approximation, which represents the 
photon absorption process by the hadronic bound state of  
the $q\bar{q}$ pair, is given by:

%% \begin{widetext}
%% \begin{multline}
\begin{eqnarray}
J^+_\pi &  = &   \nonumber 
 e (p^{+}+p^{\prime +}) F_\pi(q^2)  \\  \nonumber
 & = & \imath e
\frac{m^2}{f^2_\pi} 
N_c \int 
\frac{dk^-dk^+d^2k_{\perp}} {2 (2\pi)^4}
\frac{ Tr[ {\cal O}^{+} ]\Gamma(k,p^{\prime})
\Gamma(k,p)} 
{k^+ (k^- - \frac{f_1-\imath \epsilon}{k^+})} 
\\ 
& \times &
\biggl[ 
\frac{1}{(p^+ - k^+)(p^{-}-k^- - 
\frac{f_2 -\imath \epsilon }{p^{+} - k^+})
}
\biggr]
%%  \\   %% \nonumber \\ & &  
%% \times 
\biggl[
\frac{1}{  
(p^{\prime+} - k^+)(p^{\prime-}-k^- - \frac{f_3-\imath \epsilon} {p^{\prime+} - k^+}) } 
\biggr] , 
\label{jpion}
%% \end{multline}
 \end{eqnarray}
%% \end{widetext}
%%
where $f_i~(i=1,2,3)$ functions above are defined 
by~$f_1=k_{\perp}^2+m^2$,~$f_2=(p-k)_{\perp}^2+m^2$ and 
$f_3=(p^{\prime}-k)_{\perp}^2+m^2$, and ~the light-front coordinates by ~$a^{\pm}=a^0 \pm a^3$ and 
$\vec{a}_{\perp}=(a_x,a_y)$~\cite{Brodsky1998,Harindranath2000}.

In the electromagnetic current expression,~Eq.(\ref{jpion}), the 
Jacobian for the light-front coordinates transformation is 
$1/2$, and the Dirac trace, for 
the operator $\cal{O}^{+}$, is written (in light-front coordinates) within 
the Breit frame with Drell-Yan condition~($q^+=0$) as

%%
%% \begin{multline}
%% \begin{aligned}
\begin{equation*}
	\operatorname{Tr}
	\left[\mathcal{O}^{+}\right]=
\biggl[	 - 4 k^{-}\left(P^{\prime+}-k^{+}\right)
	\left(P^{+}-k^{+}\right)+ 
	\\ 4 \left(k_{\perp}^{2}+m^{2}\right)
	\left(k^{+}-P^{+}-P^{\prime+}\right) \\
	- 2 \vec{k}_{\perp} \cdot\left(\vec{P}_{\perp}^{\prime}-\vec{P}_{\perp}\right)
	\left(P^{\prime+}-P^{+}\right) + k^{+} q_{\perp}^{2}
\biggr]~.
\nonumber
\end{equation*}
%% \end{aligned}
%% \end{multline}   

In this case (Breit frame and Drell-Yan condition, with $\alpha=0$), we have the 
following result:

 \begin{eqnarray*}
  Tr[ {\cal O }^{+}  ]=  [-4 k^- (k^+-p^+)^2 +
   4 (k^2_\perp+m^2) (k^+-2 p^+) + k^+ q^2]~.
 \end{eqnarray*}

The quadri-momentum integration of Eq.~(\ref{jpion})  has two contribution intervals: 
(i) $0<k^+<p^+$ and (ii) $p^+<k^+<p^{\prime +}$, where $p^{\prime +}=p^+ + \delta^+$.
The first interval,~(i), is the contribution to the valence wave function for 
the electromagnetic form factor and, the second,~(ii), corresponds to the pair terms 
contribution to the matrix elements of the electromagnetic current. In the case of
non-symmetric vertex with the plus component of the electromagnetic current, 
the second interval does not give any contribution to the current matrix elements, 
because the non-valence terms' contribution is zero~\cite{deMelo1999,daSilva2012}.  
This is not the case for the minus component of the electromagnetic current for the pion, 
where beyond the valence contribution we have a non-valence 
contribution~\cite{deMelo1999} for the matrix 
elements.  For the first interval integration, 
the pole contribution is 
$\bar{k}^-=\frac{f_1-\imath \epsilon}{k^+}$. 
After integrating for the light-front energy,~$k^-$, the
electromagnetic form factors with non-symmetric 
vertex and the plus component of the  electromagnetic current are 
%% 
%% \begin{multline}
\begin{eqnarray}
F^{+(i){(NSY)}}_{\pi}(q^2) 
=    \imath e 
\frac{m^2 N^2}{ 2 p^+ f^2_\pi} N_c \int \frac{%
d^{2} k_{\perp} d k^{+}}{(2\pi)^3}
\biggl[
\frac{ Tr [ {\cal O}^{+} ] } {k^+(p^{+} - 
k^+)^2
(p^{+}-k^+)^2}  \nonumber 
  \\  
   \times 
\frac{~\theta(k^+) \theta(P^+ - k^+)} 
{(p^- - \bar{k}^- - \frac{f_2 -\imath \epsilon }{p^+ - k^+})
(p^{-} - \bar{k}^- - \frac{f_3 -\imath \epsilon }{p^{+} - k^+}) 
} 
  %% \nonumber 
  ~  %% \times  
\frac{1} {
(p^- - \bar{k}^- - \frac{f_4 -\imath \epsilon }
{p^+ - k^+}) (P^{'-} - \bar{k}^- - 
\frac{f_5 -\imath \epsilon }{p^{'+} - k^+})}
\biggr], 
\label{eq5.6}
 \end{eqnarray}
%% \end{multline} 
%%\end{widetext} 

%% 
%% where the Dirac trace above (for the quark  on-shell) is:
%% \begin{eqnarray*} & & Tr[ {\cal \bar{O} }^{+} ]=   
%% [-4 \bar{k}^- (k^+-p^+)^2 + 4 (k^2_\perp+m^2) (k^+-2 p^+) + k^+ q^2].
%% \end{eqnarray*}
%%
The functions 
$f_1$,~$f_2$ and $f_3$ were already defined; the new functions above are 
~$f_4=(p-k)_{\perp}^2+m^2_R$ and
$f_5=(p^{\prime}-k)_{\perp}^2+m^2_R$. 
The light-front wave function for the pion with the non-symmetric vertex is 
\begin{equation}
\Psi^{(NSY)}(x,k_{\perp})=
\biggl[
\frac{N}{(1-x)^2 
(m_{\pi}^2-{\cal M}_0^2) (m_{\pi}^2-{\cal M}_R^2)}
\biggr] ,
\label{wavefunction}
\end{equation}
where the fraction of the carried momentum by the quark 
is $x=k^{+}/p^{+}$ and ${\cal M}_R$ function is written as 
\begin{equation}
{\cal M}_R^2={\cal M}^2(m^2,m^2_R)=
\frac{k_{\perp}^2+m^2}{x}+
\frac{(p-k)_{\perp}^2+m_R^2}{(1-x)}-p^2_{\perp} \ .
\end{equation}
In the pion wave function expression, ${\cal M}^2_0={\cal M}^2(m^2,m^2)$ 
is the free mass operator and the normalization constant $N$ is 
determined by the condition $F_{\pi}(0)=1$. 

Finally, the pion electromagnetic form factor expressed with the light-front wave 
function for the non-symmetric vertex function may be written as
%%
%% \begin{multline}
\begin{eqnarray}
F_{\pi}^{+(i)(NSY)}
(q^2)  =    \frac{m^2}{p^+ f^2_\pi} N_c 
\int \frac{d^{2} k_{\perp} d x}
{2(2 \pi)^3 x }  
\biggl[ 
-4 (\frac{f_1}{x p^+})
(x p^+ - p^+)^2 + 4 f_1
(x p^+-2 p^+) 
\nonumber
 \\  %% & & 
+ \ x p^+ q^2 \biggr] 
 \Psi^{*(NSY)}_f(x,k_{\perp}) 
\Psi^{(NSY)}_i(x,k_{\perp}) \theta(x) \theta(1-x). 
\label{form}
 \end{eqnarray}
%%  \end{multline}

In the light-front approach, besides the valence 
contribution for the electromagnetic current, there is also a contribution from the non-valence components~\cite{deMelo1999,deMelo1998v2,Naus1998}. 
The non-valence components contribution is 
calculated in the second interval of the integration~(ii), 
through the "dislocation pole method", whose development in~\cite{Naus1998}. 
The non-valence contribution to the electromagnetic form factor,  in this case, 
is given by

\begin{eqnarray}
%% \begin{multline}
 F^{+(ii)(NSY)}_{\pi}(q^2)
& = & 
\lim_{\delta^+ \rightarrow 0}  
2 \imath e \frac{m^2 N^2}{ 2 p^+ f^2_\pi} N_c 
 \int \frac{d^{2} k_{\perp} d k^{+}}{2(2 \pi)^4} 
\theta(p^+-k^+) \theta(p^{\prime +} - k^+) \nonumber 
 \\ \nonumber 
 &  &  \biggl[\frac{ Tr~[ {\cal O }^{+}] }
{k^+(p^+-k^+)^2
 (p^{+}-k^+)^2}  
\frac{1}{(p^- - \bar{k}^- - \frac{f_2 -\imath \epsilon }{p^+ - k^+})
(p^{-} - \bar{k}^- - \frac{f_3 -\imath \epsilon }{p^{+} - k^+})}
\\ 
 & &  \frac{1}{ 
(p^- - \bar{k}^- - \frac{f_4 -\imath \epsilon }
{p^+ - k^+}) (p^{'-} - \bar{k}^- - 
\frac{f_5 -\imath \epsilon }{p^{'+} - k^+})} 
\biggr] \  \propto \delta^+ = 0 ~. 
\label{zmode}
\end{eqnarray}
%%  \end{multline}

As can be seen in equation Eq.~(\ref{zmode}),~the electromagnetic form 
factor in the second integration interval is directly proportional to $\delta^+$, 
which tends to zero. Thus, in the case of non-symmetric vertex 
for the plus component
 of the electromagnetic current, calculated for Breit frame and 
 the Drell-Yan condition~$(q^+=0)$, the non-valence (or the pair terms)
  contributions for the pion electromagnetic form factor is zero ~\cite{deMelo1999}.

For the minus component of the electromagnetic current,~$J^{-}_{\pi}~(=J^0-J^3$), 
it  is possible to extract the pion electromagnetic 
form factor with the non-symmetric vertex~(Eq.~(\ref{nosymm})). 
In this case, we have two contributions: the valence,  
for the wave function, 
and the non-valence, to the electromagnetic matrix elements of 
the electromagnetic current~\cite{deMelo1999,deMelo2002,Naus1998}. 
The pion electromagnetic form factor for the minus component 
of the electromagnetic current,~$J_{\pi}^{-}$, is related to the Dirac 
matrix by $\gamma^{-}=\gamma^{0}-\gamma^{3}$, as 
known from the light-front approach~\cite{Brodsky1998,Harindranath2000}.  
With the non-symmetric vertex,  the minus
component of the electromagnetic current is given by

%% \begin{multline}
%%\begin{widetext} 
\begin{eqnarray}
J^{-(NSY)}_\pi  & = &
e (p+p^{\prime})^{-} F^{-(NSY)}_\pi(q^2) 
%% \nonumber \\
=   \imath  e^2
\frac{m^2}{f^2_\pi} 
N_c\int \frac{d^4k}{(2\pi)^4} 
Tr \biggl[ \frac{\psla{k}+m}{k^2-m^2 + \imath \epsilon } 
\nonumber \\ 
& & \gamma^5
 \frac{\psla{k}-\psla{p'}+m}{(p'-k)^2-m^2+ \imath \epsilon}
\gamma^{-}   
 %% \left.
  \frac{\psla{k}-\psla{p}+m} {(p-k)^2-m^2+\imath \epsilon}  
\gamma^5 \Gamma(k,p^{\prime}) \Gamma(k,p)
\biggr] .
\label{j-pion}
\end{eqnarray} 
%%\end{widetext}
%%  \end{multline}

The Dirac trace in Eq.~(\ref{j-pion}), calculated with the light-front approach, results 
in the following expression:

\begin{eqnarray}
Tr[ {\cal O }^{-}] & = & \bigl[ -4 k^{-2} k^{+}     
- 4 p^{+} ( 2 k^2_{\perp} + k^+ p^+ + 2 m^2)  + k^{-} (4 k_{\perp}^{2} 
+ 8 k^+ p^+ +q^+ + 4 m^2) \bigr].
\label{tracejm}
\end{eqnarray}

The expression above has terms proportional to  
$k^{-2}$ and $k^-$. We show below that 
these terms do not disappear in the second interval integration 
and contribute to the matrix elements of the minus
component of the electromagnetic current.

In order to calculate the pair terms contribution for the minus 
component of the electromagnetic current in the 
second interval integration,~($p^+ < k^+ < p^{\prime +}$), the $k^{-}$ 
dependence in the trace and the matrix element of the pair terms are written as

%% 
%%  \begin{multline}
%% \begin{widetext}
\begin{eqnarray}
\hspace{-0.9cm}
& & J^{-(ii)\;(NSY)}  =    
\lim_{\delta^+ \rightarrow 0}  2 \imath e \frac{m^2 }{ f^2_\pi} N_c 
\int \frac{ d^{2} k_{\perp} d k^{+}}{2(2 \pi)^4} 
\theta(p^+-k^+)\theta(p^{\prime +} - k^+)
\nonumber  \\
&  &   
\biggl[ \frac{Tr[ {\cal \bar{O} }^{-}] } {k^+ (p^+ - k^+) 
(p^{\prime +} -k^+) 
}    
%% \\  & &   %%   \times
\frac{ 1}{ 
(\bar{k^-}-\frac{f_1 - \imath \epsilon}{k^+}) 
(p^- - \bar{k}^- - \frac{f_2 -\imath \epsilon }{p^+ - k^+})
} 
\frac{1}{
(p^- - \bar{k}^- - \frac{f_4 -\imath \epsilon }
{p^+ - k^+}) (p^{'-} - \bar{k}^- - 
\frac{f_5 -\imath \epsilon }{p^{'+} - k^+})
} 
\biggr], 
\label{jmin}
 \end{eqnarray}
%% \end{widetext} \end{multline}
%% 
where $p^{\prime +}=p^+ + \delta^+$ and 
$\bar{k}^{-}=p^- -
\frac{f_3-\imath \epsilon}{p^{\prime+}-k^+}$. 
The pair terms contribution for the minus component of the 
electromagnetic current is obtained with~Eq.~(\ref{jmin}), and the 
Breit frame is recovered in the limit $\delta^+ \rightarrow 0$,
%% 
%%
%% \begin{multline}
%% \begin{widetext}
 \begin{eqnarray}
J_{\pi}^{-(ii)\ (NSY)}  =  
4 \pi \biggl( \frac{m_{\pi}^2+q^2/4}{p^+} \biggr) 
\int \frac{d^2 k_{\perp}}{2 (2 \pi)^3}
%%   \times 
\sum_{i=2}^{5}\frac{ \ln(f_{i})}
{\prod_{j=2,i\neq j}^{5}(-f_i + f_j)}.
\label{jmin2} 
\end{eqnarray}   
%%\end{widetext}
%% \end{multline}
%%

This last equation, 
Eq.(\ref{jmin2}), for the minus component of the current with the 
second interval integration, is not zero and contribute to the 
electromagnetic current. This contribution is the non-valence terms contribution to the 
matrix elements of the electromagnetic current. 

The pion electromagnetic form factor with the non-valence contribution, built 
with the minus component of the matrix elements of the electromagnetic current 
calculated in Eq.(\ref{jmin2}), has the final structure 

%%
%% \begin{multline}
\begin{eqnarray}
F_{\pi}^{-(ii)\ (NSY)}(q^2)  =  
\frac{N^2}{2 p^{-}} \frac{m^2}{f^2_{\pi}} N_c
\biggl( 4 \pi \frac{m_{\pi}^2+q^2/4}{p^+} \biggr) 
%%% \\  \times 
\int \frac{d^2 k_{\perp}}{2 (2 \pi)^3}\sum_{i=2}^{5}\frac{ \ln(f_{i})}
{\prod_{j=2,i\neq j}^{5}(-f_i + f_j)}. 
\end{eqnarray}
%%  \end{multline}
%% 

The full electromagnetic form factor of the pion, for the minus component of the 
electromagnetic current, is then the sum of the partial 
form factors~$F_{\pi}^{-(i)}$~and $F_{\pi}^{-(ii)}$, 
\begin{equation}
F_{\pi}^{-(NSY)}(q^2)=
\left[ 
F_{\pi}^{-(i)(NSY)}(q^2)+
F_{\pi}^{-(ii)(NSY)}(q^2)
\right] .
\end{equation}

If the pair terms are not taken into account, 
the rotational symmetry is broken and 
the covariance is lost for the
$J_{\pi}^{-}$ component of the electromagnetic current, 
as can be seen in Fig.~1. After we add the pair terms or zero modes
contribution to the 
calculation of the electromagnetic form factor with the minus
component of the 
electromagnetic current, the identity 
\begin{equation}
F_{\pi}^{-(NSY)}(q^2)=F_{\pi}^{+(NSY)}(q^2)~
\end{equation}
is obtained  and the full covariance is restored.

In the following step, it is employed the symmetric 
vertex~$\pi-q\bar{q}$ with the plus component, "+", 
of the electromagnetic current~(Eq.~(\ref{symm})), as applied in ~\cite{deMelo2002}. 
This vertex is symmetric by the exchange of the quadri-momentum
of the quark and the anti-quark. In the light-front coordinates it is written as 

%% 
%% \begin{multline}
\begin{eqnarray} 
\Gamma(k,p)  = 
{\cal N}  \left[k^+\left(k^{-} -
\frac{k^2_{\perp}+m^2_R -\imath\epsilon}{k^+}
\right)  \right]^{-1}
+   %%   \\
 {\cal N} \left[(p^+ - k^+)
\left(p^- - k^- - \frac{(p-k)^2_{\perp}+m^2_{R}-\imath\epsilon}
{p^+ - k^+} \right)
\right]^{-1}.
\label{syvertex}
\end{eqnarray}
%%  \end{multline}
%%

With the symmetric vertex, the pion valence wave function 
results in the expression
                          
%%
%% \begin{widetext}
%%  \begin{multline}
 \begin{eqnarray} 
\Psi^{(SY)}(x,\vec k_\perp)  =  
  %% \nonumber 
\biggl[\frac{{\cal N}}
{(1-x)(m^2_{\pi}-{\cal M}^2(m^2, m_R^2))} 
+ 
%%  \\
\frac{{\cal N}}
{x(m^2_{\pi}-{\cal M}^2(m^2_R, m^2))}
\biggr]
\frac{p^+}{m^2_\pi-M^2_{0}}.
\label{wf2}
 \end{eqnarray}
%%  \end{multline}
%% \end{widetext}
%%

The electromagnetic form factor for the pion valence wave function
 (above equation) is calculated within the Breit frame, i.e.,~$(q^{+}=0)$,

%%\begin{widetext}
\begin{eqnarray}
F_\pi^{(SY)}(q^2) & = &  
\frac{m^2 N_c}{p^+f_{\pi}^2}
\int \frac{ d^{2} k_{\perp}}{ 2(2\pi)^3 }
\int_0^{1} \- \- dx
%% \nonumber \\
\left[ k_{on}^- p^{+ 2} +   
\frac14 x p^{+} q^2 
\right ]  \nonumber
\times 
 \frac{\Psi^{*(SY)}_{f}(x,k_\perp)\Psi^{(SY)}_{i}(x,k_\perp)}{x (1-x)^2},
\end{eqnarray}
%% \end{widetext}
where the on-shell condition for the 
spectator quark is $k_{on}^-=(k_{\perp}^2+m^2)/k^+$ and the normalization
constant ${\cal N}$ is determined from the condition $F^{(SY)}_\pi(0)=1$. 

The pion electromagnetic form factor calculated 
with the symmetric wave function is presented in Fig. \ref{Fig5},  
for higher momentum, and for low momentum transfer.  
In both regions, the differences between the symmetric and 
non-symmetric vertex are not very large. 

The pion decay constant, measured in the weak leptonic decay,  
with partial axial current conservation given by
\linebreak  
$P_{\mu}<0|\bar{q} \gamma^\mu \gamma^5 \tau_i q/2|\pi_j> = 
\imath m^2_{\pi}\delta_{ij} $~\cite{deMelo1999,deMelo2002} 
and vertex function $\Gamma(k,p)$, is expressed by  
\begin{equation}
\imath f_{\pi}P^{2}  = N_{c} \frac{m}{f_{\pi}} \int\dfrac{d^4k}{(2\pi)^{4}}
 Tr \left[\psla{p} \gamma^{5} S(k) \gamma^5 S(k-p)  \right] 
  \Gamma(k,p) .
\label{decay1}
\end{equation}
In the case of the non-symmetric and symmetric vertices, 
(see Eqs.~(\ref{nosymm}) and (\ref{symm})),
the expressions for the decay constant are, respectively, 
\begin{eqnarray*}
	f^{(NSY)}_{(\pi)}=\dfrac{m^2N_c }{f_\pi }
	\int\dfrac{d^{2}k_\perp}{4\pi^3}~\dfrac{dx}
	{ x }
	\Psi_{\pi}^{(NSY)} (k^+,\vec k_{\perp};m,\vec{0})~,
\end{eqnarray*}
and, 
\begin{eqnarray}
f^{(SY)}_{(\pi)}~=~ \dfrac{m^2 N_c}{f_\pi }
\int \dfrac{d^2k_\perp dx }{4 \pi^3 x (1-x)}
\Psi_{\pi}^{(SY)} (x,\vec k_{\perp};m,\vec{0})~.
\end{eqnarray}

The obtained values  of the decay constant
with the expressions above, for both light-front models calculations, 
do not have significant discrepancies, and agree with the experimental value~\cite{PDG}.

In the next section, the vector meson dominance model~(VMD)
 is presented; subsequently, we compare the VMD with the models presented here so far.

\section{Vector Meson Dominance} 

In the 1960's, J. Sakurai~\cite{Sakurai1960,Feynman1973} proposed the theory of 
{\it Vector Meson Dominance} (VMD): 
a theory of strong interactions with the local gauge invariance,
mediated by vector mesons and based on the non-Abelian field theory 
of Yang-Mills. It is possible to have two Lagrangian formulations of the 
vector meson dominance. The first was introduced by 
Kroll, Lee and Zumino~\cite{Kroll1967} and is customarily called 
VMD-1. The pion 
electromagnetic form factor calculated with this formulation results in 
\begin{equation}
F^{VMD1}_{\pi}(q^2)=
\left[ 1-
\frac{q^2}{q^2 - m^2_{\rho}}
\frac{g_{\rho \pi \pi }}{g_{\rho}}  
\right] \; .
\label{vmd1}
\end{equation}

The equation above for the electromagnetic form factor satisfies the condition $F_{\pi}(0)=1$, 
independently of any assumption about the 
coupling constants,~$g_{\rho \pi \pi}$ and $g_{\rho}$.

In the second formulation of the vector meson dominance, 
the Lagrangian has a photon mass term, and the photon 
propagator has a non-zero mass; this version is usually called~VMD-2.
With this second formulation of the vector meson dominance, 
the pion electromagnetic form factor is written as
\begin{equation}
F^{VMD2}_{\pi}(q^2)=
\left[-
\frac{m^2_{\rho}}{q^2 - m^2_{\rho}}
\frac{g_{\rho \pi \pi }}{g_{\rho}}  
\right] \; .
\label{vmd2}
\end{equation}

In the equation above, the condition $F_{\pi}(0)=1$ must be satisfied only if the 
universality limit is taken into account or translate into the following 
equality: ~$g_{\rho \pi \pi} ~=~g_{\rho}$. 
In the universality limit, as advocated by J.~Sakurai, the two formulations of the 
vector meson dominance are equivalent. 
For the present work, in Eq.~(\ref{vmd1}) and Eq.~(\ref{vmd2}), 
the rho meson mass input is the experimental value, 
$m_\rho=0.767$~GeV~\cite{PDG}, and, from the universality, $g_{\rho \pi \pi}=g_{\rho}$, 
the results at zero momentum for both equations satisfy $F_{\pi}(0)=1$.

In the present case here,  only the 
lightest vector resonance rho meson is taken into account in the 
monopole model of the VMD expressed in Eq.~(\ref{vmd1}) or 
Eq.(\ref{vmd2}). 
The vector meson dominance works quite well in the
time-like region below the $\pi\pi$ threshold.
At low energies, for the space-like region, 
the vector meson dominance model provides a reasonable description for the 
pion electromagnetic form factor. For more details and results about 
the vector meson dominance, see~\cite{Krein1993,deMelo2005,deMelo2006}.

\section{Results}   

The pion electromagnetic form factor, 
presented consistently with previous works, is extended at higher momentum 
transfer region, for $Q^2=-q^2$ up to 20~(GeV/c)$^2$~(see 
the Fig.~\ref{Fig6}). The models of the $\pi-q\bar{q}$ vertices, 
i.e, non-symmetric and symmetric vertices~\cite{deMelo1999,deMelo2002}, 
are compared with the vector meson dominance (VMD),  and   are show in Figs. \ref{Fig5} and \ref{Fig6}, 
for low and higher momentum transfer. 
%%

%% {\color{red}
	Important consequences are the direct results of the idea that
	hadrons are a composite system. Such consequences are associated
	with deep inelastic scattering (DIS), at high transferred moments.
	The applications of perturbative QCD (pQCD), at transferred high moments, 
	is certainly an extremely interesting physics problem. The pQCD 
	was initially developed in the works of Brodsky and Lepage, among others
	\cite{Lepage1980,Farrar1979,Efremov1979,Chernyak1980,Chang2013nia},
	 predicts a limit for
	$Q^2 >> \Lambda_{QCD}\sim 200~MeV$, given
	by the expression below,
	\begin{equation}
	Q^2 F_\pi(Q^2)|_{Q^2 \rightarrow \infty} \longrightarrow 16 \alpha_s(Q^2) f^2_{\pi},  
	\end{equation}
	with $\alpha_s(Q^2)$ is expressed by,    
	
	$$\alpha_s(Q^2)=\frac{4 \pi}{ \beta_0 \ln(Q^2/\Lambda_{QCD})}, 
	\ \ \ \mbox{here} ,  \ \ \ \beta_0=11-\frac{2}{3} n_f, $$
	where, $n_f$ is the flavors number.
	The two models used in the present work, 
	at moments transferred above $50.0~GeV^2$, are very close to the results obtained with the  pQCD.
%% 	} 

The pion electromagnetic radius is calculated with the derivative of the 
electromagnetic form factor for the 
pion,

~$<r^2>=-6 dF(q^2)/dq^2_{|q^2 \simeq 0}$, for both vertex
 models presented here.

In the case of the non-symmetric vertex, 
the pion radius is used to fix the parameters of the
model. The parameters are the quark mass 
$m_q=0.220$~GeV and the regulator mass~$m_R=1.0$~GeV. The pion mass used as input is the experimental
value, $m_{\pi}=0.140$~GeV. The experimental radius of the pion is
$r_{exp}=0.672\pm0.02$~fm~\cite{Amendolia1984,PDG}. 
Using the pion decay constant calculation in the non-symmetric vertex model
and with the parameters above,  the 
pion decay constant obtained is~$f_{\pi}=92.13$~MeV, 
which is close to the experimental value,~$f_{\pi} \simeq 92.28(7)$~\cite{PDG}.

On the other hand, for the symmetric vertex, the parameters are the quark mass~$m_{q}=0.220$~GeV, 
the regulator mass~$m_{R}=0.600$~GeV and the experimental mass
of the pion~$m_{\pi}= 0.140$~GeV. Our choice for the regulator mass fits the experimental 
pion  decay constant quite well when compared to the experimental data~\cite{PDG}.
 Also, good results are obtained for the electromagnetic pion radius for both vertex models. 
Both light-front models, with symmetric and non-symmetric vertices, 
are in good agreement with the experimental data 
at low energy; however, 
some differences are noticeable in the $Q^2 \geq$~1.0 (GeV/c)$^2$ region  
(see Fig.\ref{Fig5}).

The experimental data collected from  \cite{Baldini1999} worked 
well up to 10~(GeV/c)$^2$~for both the symmetric and non-symmetric vertex functions. 
For the minus component of the electromagnetic current,~$J^-$,~the pair terms or non-valence components 
of the electromagnetic current contributions are essential to
 obtain the full covariant pion electromagnetic form factor while respecting the covariance.

\begin{center}
	\begin{table}[tbh]
		\centering
		\caption{Results for the low-energy electromagnetic
			$\pi$-meson observables with the light-front models presented here.
			Immediately below are values from other models in the literature.}
		\vspace{0.20cm}
		%% \begin{indented}
		\item[]
		\begin{tabular}{l|l|l|l}
			\hline
			Model &                                $r_{\pi}$ (fm) & $f_{\pi}$ (MeV) &  $r_{\pi}.f_{\pi}$ \\
			\hline
			Non-Sym. Vert.(LF)                       &~0.672       &~93.13     &~0.316 \\
			Sym.Vert. (LF)                           &~0.736       &~92.40     &~0.345  \\
			Kissilinger et al.~\cite{Kisslinger2001} &~0.651       &~91.91     &~0.304   \\
			Silva et al.~\cite{daSilva2012}          &~0.672       &~101.0     & ~0.343     \\
			Maris \& Tandy~\cite{Maris:2000sk}       & ~0.671      & ~92.62    &~0.315     \\  
			Faessler et al.~\cite{Faessler2003}      & ~0.65       &~92.62     &~0.304      \\  
			Ebert et al.~\cite{ebert2006,ebert2005}  &~0.66        &~109.60    &~0.367       \\
			Bashir et al.~\cite{Bashir:2012fs}       &~            &~101.0       &             \\  
			Chen \& Chang~\cite{Chen:2019otg}        & ~            &~93.0        &              \\
			Hutauruk et al.~\cite{Hutauruk:2016sug}  & ~0.629       &~93.0      &~0.308              \\
			Ivanov et al.~\cite{Ivanov:2019nqd}      & ~            &~92.14     &                \\
			Jia \& Vary~\cite{Jia:2018ary}           &~0.68(5)      & ~142.8    &~0.491           \\
			Maris \& Roberts~\cite{Maris1998}        &~0.550        &~92.0      &~0.256                  \\
			Chang et al.~\cite{Chang2013nia}         &~0.66         &~92.2      &~0.307                  \\
			Eichmann~\cite{eichmann2019single}       &              &~92.4(2)   &~                  \\
			Miramontes et al.~\cite{lopez2021elucidating}   & ~0.685~($\eta=1.5$)    &~97.57     &~0.339                  \\
			Miramontes et al.~\cite{lopez2021elucidating}   &   ~0.683~($\eta=1.6$)    &~97.57     &~0.338                  \\
			Dominguez et al.~\cite{Dominguez2007}      &~0.631       & ~~       & ~                       \\
			Exp.~\cite{PDG}                        &~$0.672(8)$   &~92.28(7) & 0.313                    \\  
			\hline
			%%\protect
		\end{tabular}
		\label{table1}
		%%  \end{indented}
	\end{table}
\end{center}

In Table \ref{table2} we show the non-valence contribution
to the electromagnetic current calculated with the 
light-front approach for some values of the momentum transfer, for both instant
form and light-front approaches. 
It is worth noting that as there are no contributions due
 to valence terms or zero modes, results for the covariant calculations with both components 
of the electromagnetic
current, i.e, plus and minus 
components,~$J^{+(Cov).}, J^{-(Cov.)}$, are exactly the same.

One can also see the need to add zero modes to 
the electromagnetic current matrix elements for the minus component of
 the electromagnetic current case. 
The constituents quark models formulated with the light-front approach
 presented here are in good agreement with experimental data~\cite{Amendolia1984,Amendolia1986,Baldini1999,Volmer2001,Horn2006,Tadevosyan2007}.  

\begin{figure}[htb]
 	\begin{center}
 		\epsfig{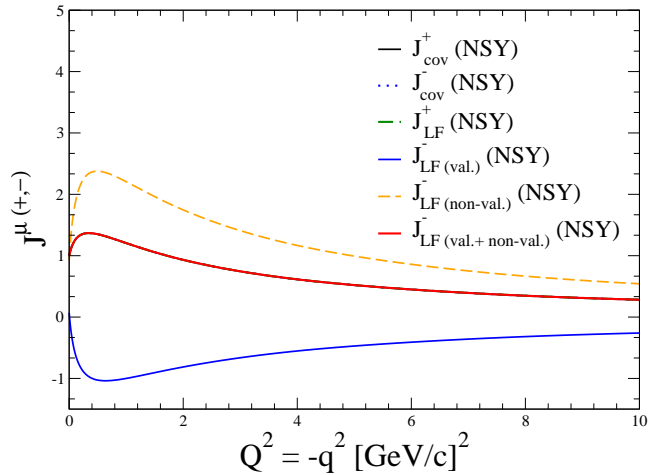} 
 	%%	. \vspace{0.50cm}
    %%	\epsfig{figure=jcurrentsy45v1.eps,width=7.50cm,angle=0}  
 		%%
\caption{The electromagetic current for the non-symmetric vertex. 
The range for the momentum transfer given here is up to 10~(GeV/c)$^2$.
}
\label{Fig2}  %% jcurrnosy}  
 	\end{center} 
 \end{figure}

The results presented in Table \ref{table2} can be visualized in Fig. \ref{Fig2}, 
where we show, for both components of the electromagnetic current,
 the calculations with both formalism: the covariant and the light-front 
approaches.

\begin{figure*}[thb]
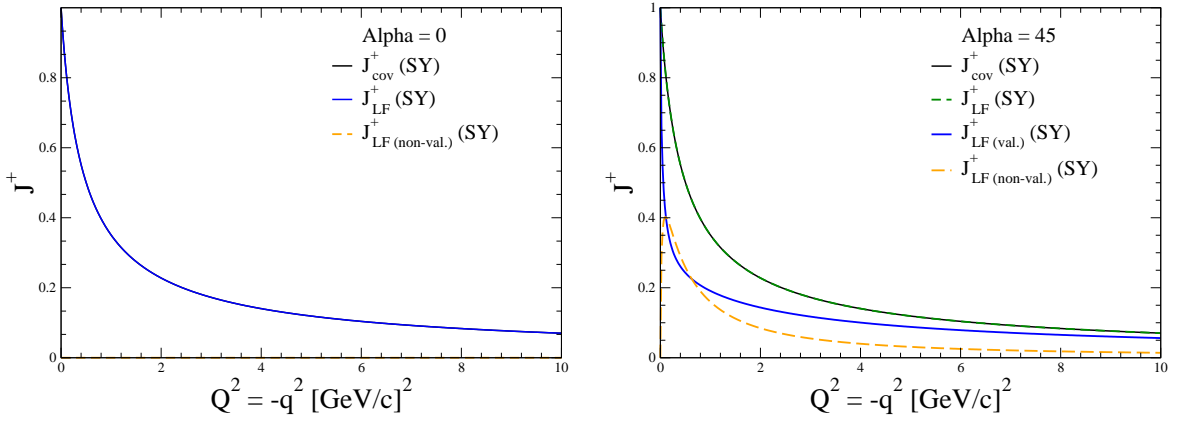

	\begin{center}
		\epsfig{figure=fig3acurr.eps,width=7.50cm,angle=0}  %% Fig3a =jcurrentsy0v1.eps
\hspace{0.25cm}	
	%%. \vspace{1.0050cm} 	 \\
		\epsfig{figure=fig3bcurr.eps,width=7.50cm,angle=0}  %% Fig3b = jcurrentsy45v1.eps
		\caption{The electromagetic current for the symmetric vertex,
			with referencial angle 0.0 (left panel) and 45 degrees
			(right panel). 
			   The range for the 
			momentum transfer given here is up to 10~(GeV/c)$^2$.
		}
		\label{jsyv0} %% alpha = 0.0 
	\end{center} 
\end{figure*}

The plus component,~$J^+$, of the 
electromagnetic current for the symmetric vertex case,
considered in this work, is presented in the left panel of Fig.~\ref{jsyv0}, 
for the case where the angle $\alpha$ is equal to zero 
(and within the Breit frame with the Drell-Yan condition). 
 It can be seen that, the non-valence contributions do not contribute to the matrix elements of the electromagnetic current in this case.

\begin{center}
	\begin{table}[tbh]
		\centering
		\caption{Numerical results for the plus and minus components of the electromagnetic
			currents~( $J^+$, $J^-$), with 
			some values for the momentum transfer, for both cases of the models considered
			here:
			non-symmetric vertex model from~\cite{deMelo1999}, 
			calculated with the covariant and with the light-front approaches. 
			Labels are:  
			(I-$J^{+(Cov.)}$, II-$J^{-(Cov.)}$, III-$J^{+(NSYV)}$, IV-$J^{-(NSYV)}$, V-$J^{-(NVAL:(NSY))}$, 
			and VI-$J^{-(VAL+NVAL)}$. The momentum $Q^2$ is given in $[GeV^2].$}
		\vspace{0.20cm}
		%%%%  \begin{indented}
		\begin{tabular}{l|l|l|l|l|l|l}
			\hline 
			$Q^2$ & ~I  & ~II & ~III &  ~IV &  ~V  & ~VI \\
			\hline
			2.0    & 0.932   & 0.932  & 0.932  & -0.814  &  1.746  & 0.932 \\
			4.0    & 0.613   & 0.613  & 0.613  & -0.553  &  1.167  & 0.613  \\
			6.0    & 0.448   & 0.448  & 0.448  & -0.409  & 0.858   & 0.448   \\
			8.0    & 0.346   & 0.346  & 0.346  & -0.317  & 0.666   & 0.346    \\
			10.0   & 0.281   & 0.281  & 0.282  & -0.260  & 0.541   & 0.281    \\
			\hline
			%%\protect
		\end{tabular}
		\label{table2}
		%%  \end{indented}
	\end{table}
\end{center}

The ratios between the electromagnetic current 
in the light-front and the electromagnetic current, calculated in the 
instant form, are given by the following equations:

%% \begin{widetext}
\begin{eqnarray}  
Ra^{I} & = & \frac{J^{+(NSYV)}_{LF}}{J^{+(NSYV)}_{Cov.}}, 
\hspace{5.7cm}
 Ra^{II} = \frac{J^{-(NSYV(val.))}_{LF}}{J^{-(NSYV)}_{Cov}}, 
\nonumber \\
Ra^{III} & = & \frac{J^{-(NSYV)(val.)}_{LF(val.)}+
	J^{-(NSYV)(nval.)}_{LF}}{J^{-(NSYV)}_{Cov}}
, \hspace{2.10cm} 
 Ra^{IV}  =  \frac{J^{+(SYV)(val.+nval.)}_{LF}}{J^{+}_{Cov}}|_{(\alpha=0)}, 
 \nonumber \\
Ra^{V} & = & \frac{J^{+(SYV)(val.)}_{LF}}{J^{+}_{Cov}} |_{\alpha=90}, 
 \hspace{4.50cm} 
Ra^{VI}  =  \frac{J^{+(SYV)(val.+nval.)}_{LF}}{J^{+}_{Cov}}|_{(\alpha=90)}, 
\nonumber \\
\label{ratios}
\end{eqnarray}
%% \end{widetext}
%% 
where the non-symmetric vertex, and symmetric vertex, are utilized according 
to~Eqs.(\ref{nosymm}) and Eq.(\ref{symm}), respectively.

%%  \begin{widetext}
 %% Fig.2
 \vspace{0.85cm}
%%  \end{widetext}

%% \begin{widetext}	
%% Fig3 3a and 3b
 %% \end{widetext}

In Eq.~(\ref{ratios}), the ratio $Ra^{I}$ is the 
plus component of the electromagnetic current calculated 
in the light-front approach divided by the electromagnetic current calculated 
in the instant form; 
both calculated with the non-symmetric vertex model~\cite{deMelo2002}.
 Since the pair terms do not contribute for the plus 
component of the electromagnetic current, the ratio $Ra^I$ is 
constant~(see Fig.~\ref{Fig8}).

%% \begin{widetext}
%% Figs 4a and 4b
\begin{figure*}[thb]
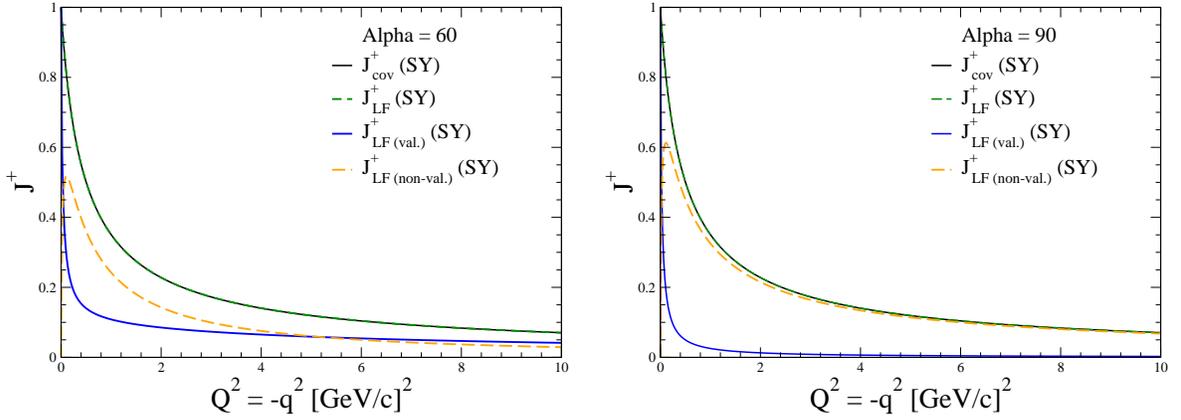

	%% 	\vspace{0.30cm}
	\begin{center}
		\epsfig{figure=fig4acurr.eps,width=7.50cm,angle=0} %% Fig.4a jcurrentsty60v1.eps
		\hspace{0.25cm}
		%%	. \vspace{1.0050cm}  \\
		\epsfig{figure=fig4bcurr.eps,width=7.50cm,angle=0} %% Fig.4b jcurrentsty90v1.eps
		\caption{The electromagetic current for the symmetric vertex,
			with referencial angle 60 (left panel) and 90 degrees (right panel). 
			The range for the 
			momentum transfer given here is up to 10~(GeV/c)$^2$.
		}
		\label{fig4} %%%diffev2}
	\end{center} 
\end{figure*}
%%\end{widetext}

The second ratio, $Ra^{II}$, is the minus component 
of the electromagnetic current, $J^{-}$, 
calculated with the light-front formalism and divided by the electromagnetic current 
calculated in the instant form. In $Ra^{III}$ ratio, the pair terms 
contribution to the 
electromagnetic current is included, so the covariance is restored. 

The ratios $Ra^{IV}$ and $Ra^{V}$ are the ``minus" components of the 
electromagnetic current without and with the pair terms 
contribution,~respectively,~divided by the ``plus" component of the electromagnetic 
current calculated in the instant form formalism.

%% \begin{widetext}
%% Fig5
   \begin{figure}
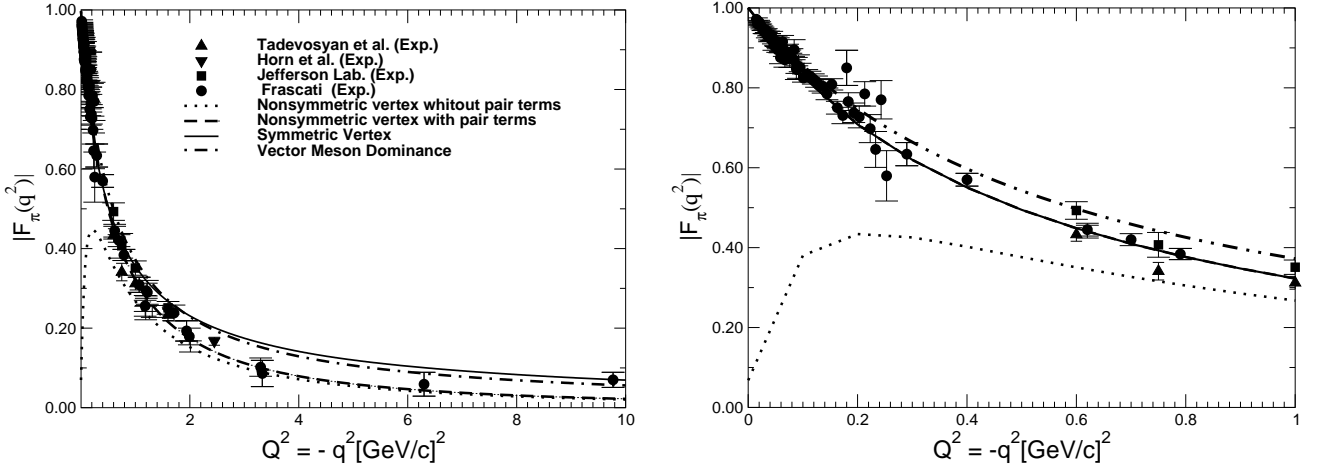
                    
  	\epsfig{figure=fig5curr.eps,width=8.4cm,angle=0}   
  	\hspace{0.25cm}
  	\epsfig{figure=fig6curr.eps,width=8.4cm,angle=0}
\caption{Left panel: Pion electromagnetic form factor calculated with light-front 
constituent quark model, for the  plus and minus components of 
electromagnetic current, compared with experimental data and vector meson 
dominance. Data are from~\cite{Volmer2001,Baldini1999,Horn2006,Tadevosyan2007}. 
Solid line is the full covariant form factor with $J^+_{\pi}$
 (symmetric vertex for the $\pi-q\bar{q}$).
 The dashed line is line the form factor 
 with $J_{\pi}^{-}$ plus pair terms
 contribution, and the dotted line is the pion form factor without  
 the pair terms contribution with the 
 minus component of the electromagnetic
 current, where both curves are with the nonsymmetric vertex. 
 After added the non-valence contribuition, the 
 pion electromagnetic form factor calculated with
 the plus or minus  compoenent of the electromagnetic current give the
 same results for the nonsymmetric  vertex.
Right panel: Pion electromagnetic form factor for small $Q^2$.
~Labels are the same as the left panel.
 %% those in Fig.~\ref{Fig5}
}
\label{Fig5}
\end{figure}
%% \end{widetext}

%% \begin{widetext}
%% Fig6
   \begin{figure}
   	\begin{center}
 %% \centerline{\includegraphics[width=12.0cm,height=10.0cm]
  %% {fig7curr.eps}} %%  Fig7 = fig3jpg.eps  }}
      	\epsfig{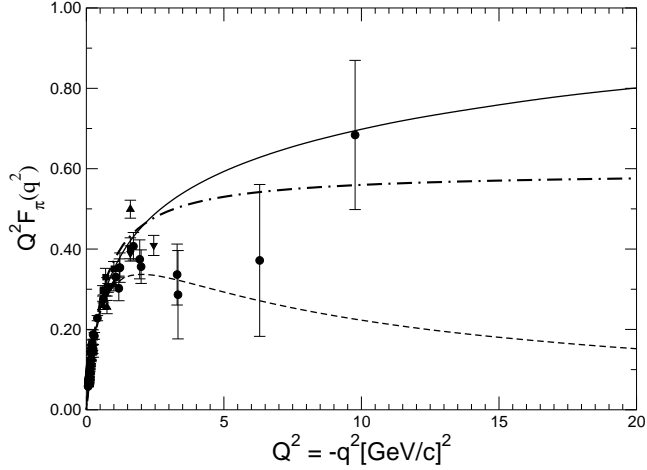}                          
\caption{Pion electromagnetic 
form factor for higher $Q^2$. Labels are the 
same as those in Fig.~\ref{Fig5}.}
\label{Fig6}  %% old Fig.3
\end{center}
               \end{figure}
%% \end{verbatim}
%% \end{widetext}

As can be seen in Fig.~\ref{Fig2}, the rotational symmetry in the 
light-front formalism is broken; this happens because the pair
 terms or non-valence contribution for the electromagnetic
  current is not taken into account properly. 
The restoration of the symmetry breaking is obtained 
by adding the 
pair terms contribution to the minus component of the electromagnetic 
current calculated in the light-front.

Experimental data for $Q^2 \gtrsim 1.5~$(GeV/c)$^2$ for the pion electromagnetic form factor
~(see Fig.~\ref{Fig5}) are not precise enough in order
 to satisfactorily select, amongst the phenomenological models,
  which is the best description for the pion elastic electromagnetic
   form factor or, in other words, the correct pion wave function.

Results displayed in Fig.~\ref{Fig6}  confirm the validity of the vector 
meson dominance model at very low momentum transfer ($Q^2\leq 0.5$~(GeV/c)$^2$).

For $Q^2>~0.5~$(GeV/c)$^2$ (see Fig.~\ref{Fig5}), however, the discrepancies between the 
vector meson dominance model, the light-front models and 
experimental data are more prominent.  
In the case of $\Delta_3$,~(see definition above), 
the covariance is respected because the difference is zero in the integration 
sum interval, [(i)+(ii)], for the $J^{-}$ component of the 
electromagnetic current. The electromagnetic form factor for the 
pion calculated with the matrix elements of the electromagnetic current
provides the same results as the
electromagnetic form factor of the pion 
calculated with usual covariant 
quantum field theory~\cite{Zuber}.

%% \begin{widetext}
%% Fig8
%%\begin{verbatim}
   \begin{figure}
%%   \centerline{\includegraphics[width=12.0cm,height=8.0cm]
%%                             {fig8curr.eps}} %% fig8 = razao21v2.eps
  \epsfig{figure=fig8curr.eps,width=7.50cm,angle=0} 
  	\epsfig{figure=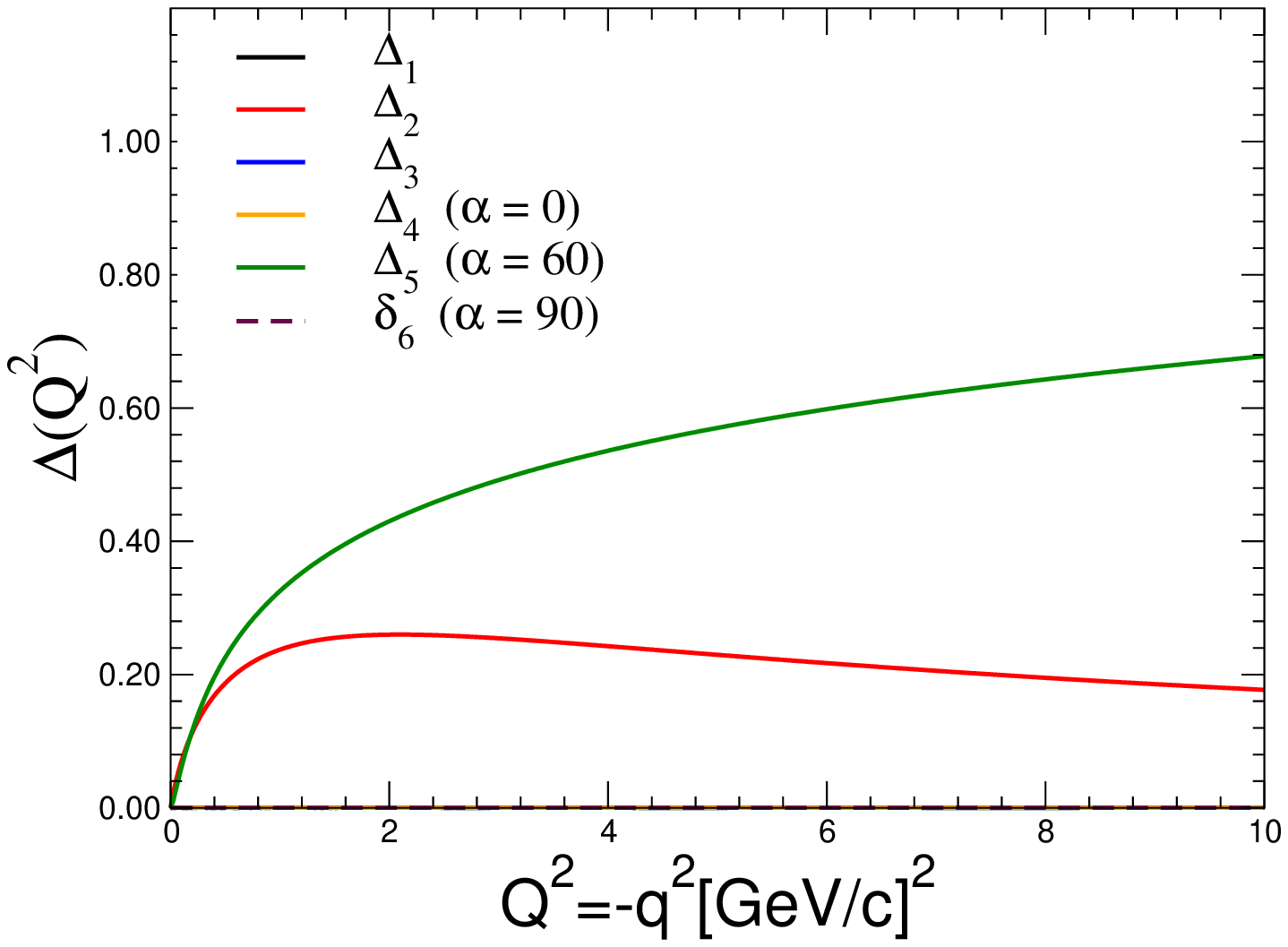,width=9.0cm,angle=0}
\caption{Left panel: Ratios for the pion electromagnetic 
current~(see Eq.(\ref{ratios}) in the text).
Right panel: Labels are the same as those in 
Eq.(\ref{diffe2}). The range for the 
momentum transfer given here is up to 10~(GeV/c)$^2$.
} 
\label{Fig8}    %%razao21}
\end{figure}
%% \end{verbatim}
%%  \end{widetext}

In order to compare the breaking magnitude of the rotational symmetry
 for the pion electromagnetic form factor with light-front models, 
 the vector meson dominance model, and the one with covariant calculations, 
 we define the following equations in an attempt to amplify the differences 
 amongst these theoretical models and experimental data:
\begin{eqnarray}
 \Delta_1   & = & \left[   
  q^2 F^{COV(NSYV))}_{\pi}(q^2) - q^2 F^{+(NSYV)}_{\pi}(q^2) \right] \; ,  
 \nonumber \\ 
\Delta_2 & = & \left[ 
  q^2 F^{(COV(NSYV))}_{\pi}(q^2) - q^2 F^{-(val.)(NSYV)}_{\pi}(q^2) \right] \; ,
\nonumber \\
\Delta_3 & = & \left[ 
  q^2 F^{(COV(NSYV))}_{\pi}(q^2) - q^2 F^{-(val.+nonval.)(NSYV)}_{\pi}(q^2) \right] \; ,
\nonumber \\
\Delta_4 & = & \left[ 
  q^2 F^{(COV(SYV))}_{\pi}(q^2) - q^2 F^{(val.)(SYV)}_{\pi}(q^2) \right]~, (\alpha=0.0) \; ,  
\nonumber \\
\Delta_5 & = & \left[ 
q^2 F^{(COV(SYV))}_{\pi}(q^2) - q^2 F^{SYV)}_{\pi}(q^2) \right]
\ , (\alpha = 60 ) \; ,  
\nonumber \\
\Delta_6 & = & \left[ 
q^2 F^{(COV(SYV))}_{\pi}(q^2) - q^2 F^{SYV)}_{\pi}(q^2) \right]
\ , (\alpha = 90 ) \; .
\label{diffe2}
\end{eqnarray}

The results of the calculations with the Eq.\ref{diffe2}, 
are shown in the 
Fig.~(\ref{Fig8})~(right panel),  up to $10~GeV^2$.

For the higher momentum transfer, 
the asymptotic behavior for the wave function of the non-symmetric vertex 
model produce  $q^2F_{\pi}(15~GeV^2/c^2)$ $\approx 0.18~$(GeV/c)$^2$. 
That result is compared  with the  leading-order-perturbative QCD, 
 $Q^2F_{\pi}(Q^2)\approx 0.15$~(GeV/c)$^2$, 
 for $\alpha_s(Q^2=10$~(GeV/c)$^2$) $\approx 0.3$ and with 
Dyson-Schwinger approach,~$Q^2F_{\pi}(Q^2)\approx 0.12-0.19~$(GeV/c)$^2$, 
for momentum transfer 
between~$Q^2~\approx ~10-15~$(GeV/c)$^2$~\cite{Maris1998}.

\section{Conclusions}

In this work, the electromagnetic form factor of the pion 
was investigated in the range $0<Q^2<20~$(GeV/c)$^2$ 
with the light-front constituent quark model. 
The light-front formalism is known nowadays as a natural way to describe 
the systems with relativistic bound state, such as the pion. %%ones.
With this approach it is possible to calculate the electromagnetic 
form factors in a more suitable way. 

However, issues related with the breaking of the 
rotational symmetry in the light-front formalism become relevant 
and the pair terms or no-valence terms contribution for the covariance restoration in 
higher energies need some attention ~\cite{deMelo1999,deMelo1997}.

After adding the non-valence components in the 
matrix elements of the electromagnetic current, the 
covariance is completely restored, and it does not matter 
which component of the electromagnetic current,~$J^+$~or~$J^-$,
~is used to extract the pion form factor with the light-front 
approach, as shown in Fig.~\ref{Fig5}. 
%% ,~\ref{Fig6} and, \ref{Fig7} %in Figs.~\ref{Fig.1},~\ref{Fig2} and \ref{Fig.3}.

The numerical results presented in~Fig.\ref{fig4} show the importance of the non-valence components for the electromagnetic current and the dependence with the choice of the component used at low momentum transfer. To achieve the full covariance, 
the inclusion of the non-valence components is essential for the minus component of the electromagnetic current. 

In Eq.(\ref{ratios}), the ratios $Ra^{I}, Ra^{III}$ and $Ra^{V}$, produce 
constant values, but the ratios 
$Ra^{II}$ and $Ra^{IV}$ do not, because the non-valence components of the electromagnetic current is not included in the light-front approach calculation~(see Fig~\ref{Fig8}).

The light-front models 
for the vertex $\pi-q\bar{q}$ and other hadronic models for the 
pion electromagnetic form-factor show good agreement between them; 
however, some differences arise when energies are 
in the higher region,~$Q^2 \gtrsim 2$~(GeV/c)$^2$.

From Eq.(\ref{diffe2}), the differences between the models analyzed 
in this work are clear, either for lower or higher momentum transfer, because the set of equations intensifies the non-similarities amongst the models. 

Since the pion electromagnetic form-factor is sensitive 
to the model adopted, it is important to compare 
different models, including new experimental data, and 
to extract new information about the sub-hadronic 
structure of the pion bound state. 

The light-front approach is a good 
framework to study the pion electromagnetic form factor.
Nonetheless, the inclusion of the non-valence components of the electromagnetic 
current is essential for both low and higher momentum transfer.

We conclude that the light-front formalism and the vertex models for  $\pi-q\bar{q}$, 
 with symmetric and non-symmetric vertices describe the new experimental data 
for the pion electromagnetic form 
factor with very good agreement. As a next step, calculations for the vector mesons, 
like $\rho$-meson and vector kaon, are in progress in
 order to allow us to also compare these with the light-front constituent models.

\vspace{0.2cm}
%%%  \section*{Ackonowlegments}

{\it Acknowledgements:}~
This work was supported in part by the Conselho Nacional de Desenvolvimento 
Cient\'{i}fico e Tecnol\'{o}gico (CNPq), Grant Process, No.~307131/2020-3 (JPBCM), 
and Funda\c{c}\~{a}o de Amparo \`{a} Pesquisa do Estado
de S\~{a}o Paulo (FAPESP), Process, No. 2019/02923-5,  
and was also part of the projects, Instituto Nacional de Ci\^{e}ncia e
Tecnologia, F\'\i sica Nuclear e Aplica\c{c}\~oes (INCT-FNA), Brazil,
Process No.~464898/2014-5, and FAPESP Tem\'{a}tico, Brazil, Process,
the thematic projects, No. 2013/26258-4 and No. 2017/05660-0.
\\ 
%%% \onecolumngrid
%% R\^omulo Moita, also, thanks the {\it Education Secretary} of the states Piau\'i and Maranh\~ao, Brazil, 
%%  for financial support.

%% The authors  declares 
%% that there is no conflict of interest regarding the publication of this paper.

%% \onecolumngrid
%% \appendix
%% references
	
%%	\clearpage  	\newpage 

%%% bbbbbbbbbbbb
%% \begin{multicols}{2}

%%%eeeeeeeee
\end{document}